\documentclass[11pt]{article}
\usepackage{amssymb}
\usepackage{makeidx}
\usepackage{graphicx}
\usepackage{array}
\def\be{\begin{equation}}
\def\ee{\end{equation}}
\def\bea{\begin{eqnarray}}
\def\eea{\end{eqnarray}}
\def\ci{\cite}

\def\bib{\bibitem}
\def\p{\phi}

\def\bib{\bibitem}

\begin{document}
\title{Cosmology and Quantum Field Theory II: Study of an extended Nambu-Jona-Lasinio model with a Dynamical Coupling }\medskip
\author{Leonardo Quintanar G. and  Axel de la Macorra \\}
\date{}
\maketitle
\begin{center}
Instituto de F\'{\i}sica,\\ Universidad Nacional Aut\'onoma de M\'exico,\\Apdo.Postal 20-364, 01000, M\'exico D.F., M\'exico
\end{center}
\vspace*{7mm}
\begin{abstract}
We study the cosmological implications of the Nambu-Jona-Lasinio (NJL model) when the coupling constant is field dependent.
The NJL model has a four-fermion interaction describing  two different phases due to quantum interaction effects and determined by the strength of the coupling constant $g$. It describes massless fermions for weak coupling  and a massive fermions and strong coupling, where a fermion condensate
is formed.  In the original NJL model the coupling constant $g$ is  indeed constant, and in this work we consider a modified version of the NJL model by introducing a dynamical field dependent coupling motivated by string theory. The effective potential as a function of the varying coupling (aimed to implement a natural phase transition) is seen to develop a negative divergence, i.e. becomes a "bottomless well" in certain limit region. Although we explain how an lower unbounded potential is not necessarily unacceptable in a cosmological context, the divergence can be removed if we consider a mass term for the coupling-like field. We found that for a proper set of parameters, the total potential obtained has two minima, one located at the origin (the trivial solution, in which the fluid associated with the fields behave like matter); and the other related to the non-trivial solution. 
This last solution has three possibilities: 1) if the minimum is positive $V_{min}>0$, the system behave as a cosmological constant, thus leading eventually to an accelerated universe; 2) if the minimized potential vanishes $V_{min}=0$, then we have matter with no acceleration ; 3) finally a negative minimum $V_{min}<0$ leads an eventually collapsing universe, \textit{even though} we have a flat geometry.Therefore, a possible interpretation as Dark Matter or Dark Energy is allowed among the behaviors implicated in the model.
\end{abstract}

\section{Introduction}

In recent times  the study of our universe has received a great deal of attention since on the one hand, fundamental theoretical questions remain unanswered and on the other hand, we have now the opportunity to measure the cosmological parameters with an extraordinary
precision.  At present time our best theoretical cosmological model contains two  unexplained forms of matter and energy called Dark Energy "DE" and Dark matter "DM" making up to $95\% $ of the energy content of the universe at present time. The existence of DE was established more then a decade ago with the study of supernovas SNIa \cite{SN.1,SN.2}, showing that the universe in not only expanding, but besides it is accelerating.

Such behavior can be explained by the existence of a new form of energy, called Dark Energy with the "strange" property of generating an anti-gravitational effect, which can be implemented by a fluid with negative pressure.
Current observational experiments involve measurement on CMB \ci{CMB} or large scale structure "LSS" \ci{LSS.1,LSS.2} or supernovae SNIa \ci{SN.1,SN.2}. Taking a flat universe dominated at present time by matter  and Dark Energy "DE", and using a constant equation of state $w$ for DE, one finds   $\Omega_{\rm DE}\simeq 0.714\pm 0.012$,  $\Omega_{\rm m}\simeq 0.286 \pm 0.012 $ with $w=-1.037^{+0.071}_{-0.070}$ from  WMAP9 results \ci{wmap9} and
$\Omega_{\rm DE}\simeq 0.6914^{+0.019}_{-0.021}$,  $\Omega_{\rm m}\simeq 0.208^{+0.019}_{-0.021} $ with $w=-1.13^{+0.13}_{-0.14}$ from  Planck \ci{planck}  and BAO \ci{BAO.1,BAO.2} measurements. The
constraint on curvature is  $-0.0013<\Omega_{k}< 0.0028$ for WMAP9 \ci{wmap9} and $|\Omega_{k}|=0.0005$ for Planck \ci{planck} using a ${\rm \Lambda CDM}$ model, i.e.  $w=-1$ for DE.

At present time, the equation of state "EoS"  of DE depends on the priors, choice of parameters  and on the data used for the analysis as can be seen from the results obtained by either WMAP or Planck collaboration groups  \ci{planck,wmap9}. A more precise determination of the EoS of DE will be carried out with e-BOSS and DESI in which together with precise measurements of CMB such as  \ci{planck,wmap9} will yield a better understanding of the dynamics of Dark Energy. With better data we should be able to study in more detail the nature of Dark Energy, a topic of major interest in the field \ci{DE.rev}. Since the properties of Dark Energy are still under investigation, different DE parametrization have been proposed to help discern on the dynamics of DE \ci{DEEoS.ax},\ci{DEparam}-\ci{quint.ax}. Some of these  DE parametrization have the advantage of having a reduced number of parameters, but they may lack a physical motivation and may also be too restrictive.  The evolution of DE background  may not be enough to distinguish between different DE models and therefore the perturbations of DE may be fundamental to differentiate between them.

On the other hand, perhaps the best physically motivated candidates for Dark Energy are scalar fields, which can interact only via gravity  \ci{SF.Peebles,tracker, quint.ax} or interact weakly  with other fluids, e.g. Interacting Dark Energy "IDE" models\ci{IDE,IDE.ax}.
Scalar fields have been widely studied in the literature \ci{SF.Peebles,tracker, quint.ax} and special interest was devoted to tracker fields \ci{tracker}, since in this case the behavior of the scalar field $\p$ is weakly dependent on the initial conditions set at an early epoch, well before matter-radiation equality. In this class of models a fundamental question is why DE is relevant now, also called the coincidence problem, and this can be understood by the insensitivity of the late time dynamics on the initial conditions of $\p$. However,  tracker fields may not give the correct phenomenology since the have a large value of $w$ at present time. However, the main objective of this phenomenological approach is to determine the dynamics of DE but does not address the question of the origin of DE.

Given that a satisfactory explanation of the nature of the two mayor contributors to the total energy of our universe, Dark Matter (DE) and Dark Energy (DE), is still missing, we think it is a good idea to use  a very interesting model of four-fermion interaction theory presented by Nambu and Jona-Lasinio \cite{NJL}, the NJL model. Other possibility is to study NJL for inflation see \cite{NJLInflation}. Alternatively, some interesting studies for DE and DM have been proposed derived from gauge groups, similar to QCD in particle physics, and have been studied to understand the nature of Dark Energy \ci{GDE.ax} and also Dark Matter \ci{GDM.ax}.

From a particle physics point of view, the NJL model is a very interesting theory involving a (chiral) symmetry breaking, the particle states before and after the breaking being related with two different physical phases. The defining lagrangian of the theory includes a single parameter, knew as the \textit{coupling} $g$, but because it is indeed a non-renormalizable theory, one must introduce a second parameter in the form of a cut-off $\Lambda$ in the energy scale, in order to regularize it. Of course, one can always consider it as a valid theory by watching that the energy keeps below the cut-off. The symmetry is broken for a "strong" coupling $g>g_c$ exceeding the critical value $g_c=2\pi/\Lambda$, in which case one has a fluid consisting of a \textit{fermion condensate}, effectively described by a scalar field. For $g<g_c$ we have a "weak" coupling leading to a fluid of massless fermions.

Some years ago, the NJL model was used to break supersymmetry in the context of superstring models \ci{NJL.ax}. The breaking of susy is a consequence of the formation of a gaugino (fermion fields) condensate which are dynamically favored. In string theory the gauge coupling constant $g_{cc}$ is field dependent, it depends on the dilaton field. So, having this in mind, we are motivated to  allow here to have a field dependent coupling $g$ in our NJL model, but our coupling $g$ is not to be confused with he gauge coupling constant $g_{cc}$.

One may consider a universe containing one or other of these NJL fluid phases, and solve the cosmological dynamics. Moreover one may consider additional "barotropic" fluids (satisfying $w_b=cte.$ in the state equation $P_b=w_b \rho_b$), including dust-matter ($w_m=0$) and radiation ($w_r=1/3$). This was done in a previous work \cite{njlcosmo}. There, it was found that a negative potential can be related to a strong coupling, and in its turn, because of the dynamical equations, it causes the scale factor eventually reaching a maximum value and further going to a contracting period forward in time, all the way to a vanishing value. In other words, a universe containing a NJL fermion condensate phase ends up in a "big crunch", \textit{even though} a flat geometry is chosen from the beginning. On the other hand, if we take reliably the effective description given in terms of the scalar field $\phi$, a weak coupling can be related to a potential function of the form $V\sim \phi^2\geq 0$ about the minimum, this implicating the NJL fluid behaves as matter with an effective state coefficient $w_{NJL}=0$. A universe containing such a fluid keeps expanding forever without acceleration. We see therefore, that the NJL model has interesting properties from the cosmological point of view as well.\\

Nevertheless, there is nothing in the theory to induce a "natural" phase transition (not put in by hand), since the original NJL model contemplates only a \textit{constant} coupling. However, we would like to investigate further the interesting features that the NJL model has to offer, and find whether these could be exploited to work towards a solution of some of the unanswered questions in cosmology. To this end, we study how a dynamical coupling could be introduced, and see if this would allow such a phase transition. This paper is organized as follows: we first summarize the main features of the model (section \ref{njlrev}). Then we argue the introduction of a dynamical coupling, and present the necessary theory in treating two scalar fields in general (section \ref{dyncoupling}). Next we proceed to study the potential obtained for our model in particular (section \ref{analysispot}). Once this is done, we can move to solve the associated cosmological dynamics (section \ref{cosmoexnjl}), and finally synthesize our conclusions (section \ref{conclu}).

\section{NJL revisited.}\label{njlrev}

In any theory, and in particular in field theory, we can consider to have a better model  if it contains the least number of parameters and if the physical values of these parameters are "natural", i.e. of order one. This was indeed one of our motivation to study the NJL model and its  cosmological consequences  in our previous  work \ci{njlcosmo}.  An inherent symmetry associated with a fermion field is the so called "chiral" symmetry, relating the "right" and "left" components of the field. We will start with the lagrangian functional  satisfying  this chiral symmetry, with the form
\begin{equation}
{\cal L}=i\bar\psi \gamma^{\mu}\partial_{\mu}\psi +\frac{g^2}{2}\left[(\bar\psi\psi)^2-(\bar\psi\gamma_5\psi)^2\right] ,
\label{Lnjl}
\end{equation}
which is precisely the model of Nambu and Jona-Lasinio with $g$  the coupling constant which is indeed constant. The field $\psi$ is a four-component spinor.  For our present purposes, we would like to take an even simpler version of the theory by neglecting the pseudoscalar contribution. In doing so, an equivalent lagrangian to eq. (\ref{Lnjl}), concerning only the interaction term, can be written in the form
\begin{equation}
{\cal L}_{int}=mg\phi\bar\psi\psi-\frac{1}{2}m^2\phi^2 ,
\label{Lnjleq}
\end{equation}
where the parameter $m$ with dimension of mass was introduced to make all physical units consistent and
the field  $\phi$, being a Lorentz invariant, plays the role of an auxiliary  scalar field. From eq.(\ref{Lnjleq}) the  equation
of motion  of the field  $\phi$ gives
\begin{equation}
\phi=\frac{g}{m}\bar\psi\psi . \label{phi}
\end{equation}
From this expression the fermion mass $m_\psi$ and the tree level scalar potential  $V_0$, are seen to be
\begin{equation}
m_\psi^2=(mg\phi)^2,\quad V_0=\frac{1}{2}m^2\phi^2 . \label{pot0}
\end{equation}
As already mentioned in our previous work \cite{njlcosmo}, the four-fermion interaction form of the theory makes it a non-renormalizable theory.
The interaction can be expanded following conventional perturbation theory, and represented by Feynman diagrams.
\begin{figure}[h]
\centering
\includegraphics[scale=.5]{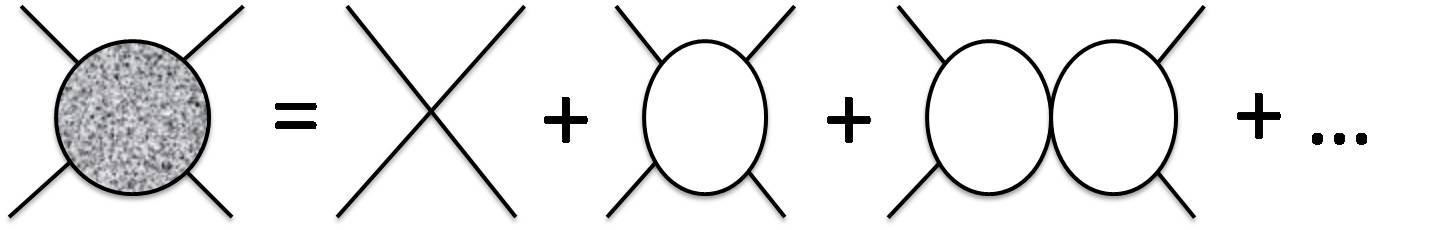}\caption{Feynman diagram for a four-fermion interaction.}
\label{fig00}
\end{figure}\\
However, the infinite number of fermion loops can be resumed giving a non-perturbative potential, which has to include a cut-off $\Lambda$ in the energy scale. The one-loop potential is given by
\begin{equation}\label{Vint}
V_1=-\frac{1}{8\pi^2}
 \int d^2p\; p^2 Log\left(p^2+m^2_\psi\right);
\end{equation}
the integral has an ultraviolet  cutoff $\Lambda$,  \ci{NJL.ax} while  the mass of the fermions $m^2_\psi$ is given by
$m^2_\psi=m^2g^2\phi^2$.
If we define
\begin{equation}
x\equiv \frac{m_\psi^2}{\Lambda^2} = {m^2g^2\phi^2\over \Lambda^2},\label{xphi}
\end{equation}
the one loop potential is given by
\begin{equation}
V_1=-\frac{\Lambda^4}{16\pi^2}
\left[
\left({mg\phi\over \Lambda}\right)^2+
\left({mg\phi\over \Lambda}\right)^4\log
\left(\frac{\left({mg\phi\over \Lambda}\right)^2}{1+\left({mg\phi\over \Lambda}\right)^2}\right)
+\log\left(1+\left({mg\phi\over \Lambda}\right)^2\right)
\right],
\end{equation}
and the  total effective scalar potential, taking the quantum corrections into account, is given by
\begin{equation}
V_I=V_0 +V_1 =\frac{\Lambda^2 x}{2g^2}-Af(x),\label{VIx}
\end{equation}
with
\begin{equation}
A\equiv \frac{\Lambda^4}{16\pi^2},\quad
 f(x)=x+x^2\log\left(\frac{x}{1+x}\right)+\log(1+x).\label{Af}
\end{equation}
Note that the coefficient "$A$" is a constant. So, the effective potential written explicitly as a function of $\phi$ is
\begin{equation}
V_I(\phi)=\frac{1}{2}m^2\phi^2 -\frac{\Lambda^4}{16\pi^2}
\left[
\left({mg\phi\over \Lambda}\right)^2+
\left({mg\phi\over \Lambda}\right)^4\log
\left(\frac{\left({mg\phi\over \Lambda}\right)^2}{1+\left({mg\phi\over \Lambda}\right)^2}\right)
+\log\left(1+\left({mg\phi\over \Lambda}\right)^2\right)
\right] .\label{vefphi}
\end{equation}
In the above expressions for the potential we have included a subindex "I" to distinguish it from additional terms to be considered later.\\
As the complete analysis and explanations of the potential were already done  in \cite{njlcosmo}, here we will limit ourselves to remember the main facts:\\
For a fixed value of the coupling $g$, such that $g<g_c$, i.e. a "weak" coupling, with a critical value
\begin{equation}
g_c= \frac{2\pi}{\Lambda},\label{gc}
\end{equation}
the potential is minimized at the origin $\phi=0$, and can be Taylor-expanded
as $V_I= {1\over 2}m^2\left( 1- {\Lambda^2 g^2\over 4\pi^2}\right)\phi^2$ about the origin. Because $g<g_c$, the coefficient multiplying $\phi^2$ is a positive quantity, so that it is $V_I\geq 0$. The dynamical evolution of the field $\phi$ is a damped oscillation about the minimum, so that the pairing $\phi\sim \bar\psi\psi$ vanishes in average, $<\phi>\sim 0$, meaning that we have a massless fermion fluid.
On the other hand, if the coupling is fixed in a "strong" value $g>g_c$, the potential is minimized in a non-trivial value $\phi=\phi_{min}$, where $\phi_{min}$ is related to $x_{min}$ through equation (\ref{xphi}), and $x_{min}$ is the solution to the transcendental equation
\begin{equation}\label{mgap}
\frac{4\pi^2}{g^2\Lambda^2}= 1+x\log\left(\frac{x}{1+x}\right),\label{eqx1}
\end{equation}
with the valuated potential at the minimum $V_{min}=V_I(\phi_{min})<0$ being a negative quantity.  Eq.(\ref{mgap}) is to the mass gap equation (see \cite{massgap})  derived from the infinite number of  fermion bubbles, shown in fig.(\ref{fig00}),  of the Schwinger-Dyson equation and a non trivial solution corresponds to a non-perturbative results since by equating a tree level result with a radiative correction one is necessarily in the non-perturbative regime and is effectively summing over the infinite number of fermion bubbles. A non trivial solution signals that a condensate is dynamically favoured. As the pairing $\phi_{min}\sim \bar\psi\psi \neq 0$ is energetically favored, in this case it is obtained a fermion condensate in the form of a scalar field $\phi$.\footnote{Observe that a solution to the equation (\ref{eqx1}) can be interpreted, from a geometrical point of view, as the intersection of the right line $4\pi^2/g^2\Lambda^2=const.$ with the curve of the function $1+x\log [x/(1+x)]$. This way of visualizing a solution will be useful to us when we search extremum points for a more complicated potential, further along.}

Now, as we are looking forward to consider a dynamical field dependent coupling, we need to know the behavior of the potential (\ref{vefphi}) as a function of two variables, $\phi$ and $g$ (or $x$ and $g$). For the function $f(x)$ defined in eq. (\ref{Af}) we have
\begin{equation}
\lim_{x\to 0}f(x)=2x, \qquad \lim_{x\to\infty}f(x)=-\log x ,\label{limf}
\end{equation}
and for the derivative
\begin{equation}
\frac{df(x)}{dx}=2\left[1+x\log\left(\frac{x}{1+x}\right)\right].\label{df}
\end{equation}
By using equations (\ref{limf}) we see that in the limit $x\sim 0$, we have $V_I\simeq \frac{\Lambda^2 x}{2}\left( \frac{1}{g^2}-\frac{\Lambda^2}{4\pi^2} \right)$ if $g\sim 0$, but $V_I\simeq -\frac{\Lambda^4}{8\pi^2}x$ if $g\to \infty$. We summarize the different limit behaviors of the potential in the following table:
\begin{center}
\begin{tabular}{|c|c||c|}
\hline\hline\hline $x$ & $g$ & $V_I(x,g)$\\ \hline\hline\hline
$x\to 0$ & $g\to 0$ & 0 (for $x<<g$),\\
 & & $\infty$ (for $x>>g$)\\  \hline
$x\to 0$ & $0<g<\infty$ & 0\\ \hline
$x\to 0$ & $g\to \infty$ & $0  (\sim -\frac{\Lambda^4}{8\pi^2}x)$\\
\hline $0<x<\infty$ & $g\to 0$ & $\infty$\\
\hline $0<x<\infty$ & $0<g<\infty$ & $V_I(x,g)$\\
\hline $0<x<\infty$ & $g\to \infty$ &$-\frac{\Lambda_\psi^4}{16\pi^2}f(x)$\\
\hline $x\to \infty$ & $g\to 0$ & $\infty$\\ \hline
\hline $x\to \infty$ & $0<g<\infty$ & $\infty$\\ \hline
\hline $x\to \infty$ & $g\to \infty$ &$\infty$ (for $\frac{x}{\log x}>>\Lambda^2g^2$),\\
 & & $-\infty$ (for $\frac{x}{\log x}<<\Lambda^2g^2$)\\
\hline \end{tabular} \end{center}
We can see from the table that in the limit situations, the potential goes to positive values, zero  or  negative  \textit{finite} values. However, there is only one situation in which the potential has not a lower bound: if $x\to \infty$, $g\to \infty$, then we can write the approximation $V_I\simeq \Lambda^2\left( \frac{x}{g^2}-\Lambda^2\log x \right)$, which under the constriction $x/\log x<<\Lambda^2g^2$, becomes
\begin{equation}
V_I\simeq -\Lambda^4\log x,\quad \mbox{i.e.}\quad V\to -\infty \quad \mbox{when }\quad x\to\infty ,\; g\to\infty .\label{limiteVI}
\end{equation}
This shows that, in considering the potential (\ref{VIx}) as a function of the coupling, there exist a path for it to have a negative divergence. This could, or could not be taken to be a serious drawback, as we now explain:\\
On the one hand, if the dynamical evolution of a field is such that it tends to minimize the potential, then one should be careful to see that such a potential has indeed a minimum. If a potential had not a lower bound, a field, in attempting to minimize the energy, would follow a path such that the potential would adopt arbitrarily lower (negative) values. In other words, the physical system would release an infinite amount of energy! This, of course, is an unacceptable result, and one should discard a potential having such a behavior, at first sight.\\
On the other hand, it was shown before in \cite{njlcosmo}, that when considering a scalar field moving under a potential having a \textit{finite negative} value $V(\phi_{min})=-|V_{min}|$ when minimized, then the cosmological dynamics dictates that the universe must eventually suffer a collapse in a finite time.\footnote{This is because, for a generic barotropic fluid the energy density $\rho_b\sim a^{-n}$ diminishes ($n>0$ and $a$ increasing in time), and for a scalar field which is stabilizing about the minimum the kinetic energy $E_k\sim 0$, $V\sim V_{min}<0$, so that the total energy density $\rho=\rho_b+E_k+V$ initiating at some \textit{positive} value, eventually goes to zero. According to the Friedman equation we have $H^2=(1/3M_p^2)\rho$, so that as time passes $H^2=(\dot{a}/a)^2\to 0$, i.e. $\dot{a}=0$; meaning the scale factor $a(t)$ reaches a maximum. Further along, because $H$ is always diminishing, it becomes $H<0$, i.e. $\dot{a}<0$, meaning $a(t)$ decreasing and eventually vanishing.} But this consequence is related to an eventually vanishing total energy density, which of course, is a situation that can not be avoided if the potential is not even limited from below. In other words, a universe containing a scalar field governed by a negatively divergent potential, is also (as well as with a negative finite potential) expected to reach a maximum size, and subsequently enter in a contracting phase, to end up in a collapse. All this takes place along a finite amount of time, so that the releasing of an infinite energy is therefore prevented. This circumstance thus allow us reconsider throwing a "bad behaving" potential (here we have an interesting result).

Although a negative unbounded potential is not necessarily a drawback (at least in the cosmological context), let us consider how we could get a bounded potential. According to (\ref{limiteVI}), the potential diminishes like $-\log x$, so if we would add a term growing faster, for instance $V_{II}\sim \log^2 x$, the total potential
$V=V_I+V_{II}\label{Vtot}$
would not go to negative values when $x\to \infty$, $g\to \infty$ anymore, because in this limit we have $V\sim -\log x+\log^2 x\to \log^2 x\to \infty$. Now, a term $\sqrt{x/\log x}$ grows faster than $\log x$, and a term proportional to $g$ would grow even faster, because $x/\log x<<\Lambda^2g^2$ was the constriction taken in eq. (\ref{limiteVI}). Thus, we see that by adding a term of the form $V_{II}\sim g^2$, the negative divergence of the potential is removed.
For the sake of clarity and simplicity, it will be convenient to work with a total potential considered as a function of a linear variable, let's say $h$, rather than of a quadratic one, $g^2$. So, let us define
\begin{equation}
h=\frac{\Lambda^2 g^2}{8\pi^2}\quad\mbox{(coupling)},\qquad V_{II}=Abh,\quad\mbox{(b constant)}\label{defh}
\end{equation}\\
where the term $V_{II}$ is written in a convenient form with a constant coefficient $b$ (whose explicit form will depend on the interpretation of the term $V_{II}$ in the potential, to be discussed further along), and the constant $A$ from eq. (\ref{Af}). It can be seen that the tree level potential in eq. (\ref{pot0}) can then be written as $V_0=Ax/h$. In this way, using the variable $h$ to rewrite eq. (\ref{VIx}), and from definitions in eq. (\ref{defh}), we have for the total potential 
as a function of $x$ and $h$:
\begin{eqnarray}
V_I=V_0 +V_1 &&= A\left[ \frac{x}{h}-f(x)\right] ,\\
V(x,h)=V_I+V_{II} &&= A\left[ \frac{x}{h}-f(x) +bh \right]. \label{Vxh}
\end{eqnarray}
Having analyzed the limit behavior of the potential, we now turn to see how a dynamical coupling can be implemented.

\section{Dynamical coupling: two scalar fields.}\label{dyncoupling}
Given any theory, a parameter can be always taken to be a function of time and see "what happens". However, one would like to have a good reason in doing so, rather than simply introducing such a varying parameter by hand, let us say, by being supported by an underlying theory. In our case, this support turns out to come from String Theory. There, it arises the possibility that a scalar field (known as the \textit{dilaton}) may play the role of a coupling constant. In fact, a variety of functional relationships are allowed, depending on the specific string theory. Because we would like to begin by studying the simpler cases, let us consider a power law functional of the general form
\begin{equation}
g=\frac{\varphi^n}{M^{1+n}},\quad\mbox{or}\quad h=\frac{\Lambda^2}{8\pi^2}\frac{\varphi^{2n}}{M^{2(n+1)}},\label{gphi}
\end{equation}
where we have introduced the mass-dimension parameter $M$ in order for the coupling $g$ (or $h$) to keep the right units. If $\varphi$ is a dynamical field, then the above equations tell us that the coupling is evolving in time.

Now, how should we take the value of $n$ in eq. (\ref{gphi})? According to conventional QFT, the simplest potential for a scalar field $\varphi$ is a quadratic potential, which is related to the mass of the field (let us name it $m_0$, a constant). A potential of this kind would be a natural choice. So, let us write $V_{II}={1\over 2}m_0^2\varphi^2$. From eq. (\ref{defh}) this means that ${1\over 2}m_0^2\varphi^2=Abh$, and substituting $h$ from eq. (\ref{gphi}), we would have
\begin{equation}
b={(8\pi)^2 m_0^2 \over \Lambda^6}  M^{2(1+n)}\varphi^{2(1-n)}. \label{defb}
\end{equation}
The only value for which the parameter $b$ is a constant (remember that this parameter was defined to be so) is for $n=1$; by choosing this value the term $V_{II}$ of the potential can be identified as being a mass term.
If so, equations (\ref{gphi}) and (\ref{defb}) then become
\begin{equation}
g=\frac{\varphi}{M^2},\quad h=\frac{\Lambda^2 \varphi^2}{8\pi^2 M^4},\quad
b=(8\pi^2)^2\frac{m_0^2 M^4}{\Lambda^6}. \label{ghb}
\end{equation}
By the way, we can look for the value of the field $\varphi$ associated with the critical value of the coupling. Using the result eq. (\ref{gc}), and the first equation (\ref{ghb}) we find
\begin{equation}
\varphi_c=2\pi\frac{M^2}{\Lambda}.\label{varphicrit}
\end{equation}
The critical value in eq. (\ref{gc}) is not affected by the addition of the term $V_{II}$ to the original potential $V_I$, because it is not a function of the another relevant field $\phi$. So the equation (\ref{varphicrit}) determine the point of the phase transition in terms of the field $\varphi$, and remains valid for the total potential eq. (\ref{Vxh}).

\subsection{A word on interacting fields.}
Given that the potential in eq. (\ref{vefphi}) include the coupling $g$ as a parameter, we could write it as a two variable function of the form $V_I=V_I(\phi,g)$. Besides, we defined a term $V_{II}=V_{II}(g)$. Now, the path we follow to consider a dynamical coupling is to introduce a second scalar field $\varphi$, by defining a functional relationship of the form $g=g(\varphi)$. So the total potential (\ref{Vxh}), $V=V_I+V_{II}$ can be seen also as  a function of two dynamical fields $V=V(\phi,\varphi)$. Using conventional field theory and conditions of spatial homogeneity, we can write the general functional lagrangian
\begin{equation}
{\cal L}=\sqrt{-g}\left[ \frac{1}{2}\dot\phi^2+\frac{1}{2}\dot\varphi^2-V(\phi,\varphi)\right] .
\end{equation}
The FRW metric has $g=det(g_{\mu\nu})=-a^6$.
The equations of motion corresponding to each field can be calculated to be
\begin{eqnarray}
\ddot\phi+3H\dot\phi+\frac{\partial V}{\partial\phi}=0,\label{ecmov1}\\
\ddot\varphi+3H\dot\varphi+\frac{\partial V}{\partial\varphi}=0,\label{ecmov2}
\end{eqnarray}
and the energy-momentum tensor can be also calculated following the standard procedure. However, in general the constituting terms of the potential are in the form of products, or functions of both fields, which can not be written as sums of functions for each individual field. At this point it arises the question of how to define the energy densities and pressures associated to each field, because the different ways in which one can group the terms are rather arbitrary.
Nevertheless, we may consider that at least one field can be separated in such a way that we can write the potential in the form
\begin{equation}
V(\phi,\varphi)=U_1(\phi)+U_2(\phi,\varphi).
\end{equation}
In this way, we can define an energy density involving a single field, and attach the entire interaction to the other field, for instance in the form
\begin{eqnarray}
\rho_{\phi}=\frac{1}{2}\dot\phi^2 +U_1(\phi),\quad &P_{\phi}=\frac{1}{2}\dot\phi^2 -U_1(\phi),\label{rhophivarphi1}\\
\rho_{\varphi}=\frac{1}{2}\dot\varphi^2 +U_2(\phi,\varphi),\quad &P_{\varphi}=\frac{1}{2}\dot\varphi^2 -U_2(\phi,\varphi). \label{rhophivarphi2}
\end{eqnarray}
A suitable choice in our case is the definition
\begin{eqnarray}
U_1(\phi)= V_0 &&= {1\over 2}m^2\phi^2 ,\\
U_2(\phi,\varphi) =V_{II}+V_1 &&= {1\over 2}m_0^2\varphi^2
-\frac{\Lambda^4}{16\pi^2}\left[x+x^2\log\left(\frac{x}{1+x}\right)+\log(1+x)\right].
\end{eqnarray}
By taking the derivative of the energy densities above, and using the equations of motion (\ref{ecmov1}, \ref{ecmov2}), we find
\begin{eqnarray}
\dot\rho_{\phi}+3H(\rho_{\phi}+P_{\phi})=\Gamma ,\label{int1}\\
\dot\rho_{\varphi}+3H(\rho_{\varphi}+P_{\varphi})=-\Gamma ,\label{int2}
\end{eqnarray}
where we have defined the \textit{interaction function} $\Gamma$ by
\begin{equation}
\Gamma=-\frac{\partial U_2}{\partial\phi}\dot\phi .
\end{equation}
Note that the energy density defined by $\rho=\rho_\phi+ \rho_\varphi$, with a pressure $P=P_ \phi+P_\varphi$, satisfies the continuity equation $\dot{\rho}+3H(\rho+P)=0$, whereas, due to the presence of the interaction function,
any individual energy density $\rho_\phi$, $\rho_\varphi$, does.

For a given barotropic fluid "$\alpha$" with an equation of state $P_\alpha=w_\alpha$, the coefficient of state $w_\alpha$ is sometimes a useful parameter to characterize the fluid. In the case of interacting fields one can define an effective coefficient of state (ECS)
\begin{equation}
({w_{ef}})_\phi=w_\phi- \frac{\Gamma}{3H\rho_\phi},\quad ({w_{ef}})_\varphi=w_\varphi +\frac{\Gamma}{3H\rho_\varphi},\label{weff}
\end{equation}
which allow us to write a "continuity" equation for the energy density of the fields in the form
\begin{equation}
\dot{\rho}_\alpha+3H\rho_\alpha[1+({w_{ef}})_\alpha]=0.
\end{equation}
A situation in which the ECS can be taken as a constant, i.e. $({w_{ef}})_\alpha\simeq cons.$, then the equation above could be solved to give the approximate solution
\begin{equation}
\rho_\alpha\propto a^{-3[1+({w_{ef}})_\alpha]},
\end{equation}
thus we find the usefulness of the ECS in helping us to see how a fluid-field is diluting.

\section{Analysis of the total potential $V=V_I+V_{II}$.}\label{analysispot}
The evolution of the fields, as well as of other cosmological variables are determined by the  potential (aside initial conditions), and in general, a wide range of different solutions are obtained depending on the specific set of parameters. Thus, we need to have some idea of the behavior of the potential in terms of the parameters involved, so we can discriminate the useful solutions matching the real observations. Although sometimes the potential may have a complicated form when expressed as an explicit function of the fields (as in our case), conventional mathematical analysis is, of course, a pertinent tool.\medskip\\

As seen at the end of section \ref{njlrev}, with the help of variables $x$, $h$, the potential adopts the more simple appearance, given in equation (\ref{Vxh}). However, we must not forget that the fields are the real relevant variables. We have $\frac{\partial V}{\partial \phi}=\frac{\partial V}{\partial x}\frac{\partial x}{\partial \phi}+\frac{\partial V}{\partial h}\frac{\partial h}{\partial \phi}$. But, according to eq. (\ref{gphi}), $h$ does not depend on $\phi$, so $\frac{\partial h}{\partial \phi}=0$. Also, according to eq. (\ref{xphi}), the derivative  ${\partial x\over \partial \phi} \sim \phi$ is linear in $\phi$, i.e. a monotonous function with single vanishing point at $\phi=0$. Therefore, the condition $\frac{\partial V}{\partial \phi}=0$ is equivalent to the following equations:
\begin{equation}
\phi=0,\quad\mbox{or}\quad
\frac{1}{A}\frac{\partial V}{\partial x} =\frac{1}{h} -2\left[ 1+x\log\left(\frac{x}{1+x}\right) \right].\label{dVx}
\end{equation}
Similarly, for the derivative w.r.t. $\varphi$ we have $\frac{\partial V}{\partial \varphi}=\frac{\partial V}{\partial x}\frac{\partial x}{\partial \varphi}+\frac{\partial V}{\partial h}\frac{\partial h}{\partial \varphi}$. From the second equation in the definitions (\ref{gphi}) we have that $h$ is a power law function in $\varphi$, so that ${\partial h\over \partial \varphi}\sim \varphi^{n-1}$ is a monotonous function vanishing at $\varphi=0$. From definitions (\ref{xphi}) and (\ref{gphi}) for $n=1$, we can see that it is ${\partial x\over \partial \varphi}\sim \varphi$, so that the condition $\frac{\partial V}{\partial \varphi}=0$ has $\varphi=0$ as a solution. Because an extremum point must satisfy the conditions $\frac{\partial V}{\partial \phi}=0$, and $\frac{\partial V}{\partial \varphi}=0$ \textit{simultaneously}, we then see that the origin $\phi=0, \;\varphi=0$ is an extremum point.
The second order derivatives valuated on the origin gives:
\begin{equation}
D_1=\left.{\partial^2 V\over \partial\varphi^2}\right\vert_{(0,0)} =m_0^2,\quad
D_2=\left.{\partial^2 V\over \partial\phi^2}\right\vert_{(0,0)} =m^2,\quad
D_3=\left.{\partial^2 V\over \partial\varphi \partial\phi}\right\vert_{(0,0)} =0,
\label{segderivada}
\end{equation}
with the determinant $D_1D_2-D_3^2=m_0^2m^2>0$ being a positive quantity.
So the standard mathematical criterion tell us that the extremum (in this case the origin) is indeed a local minimum.

\begin{figure}[h] \centering \includegraphics[scale=.5]{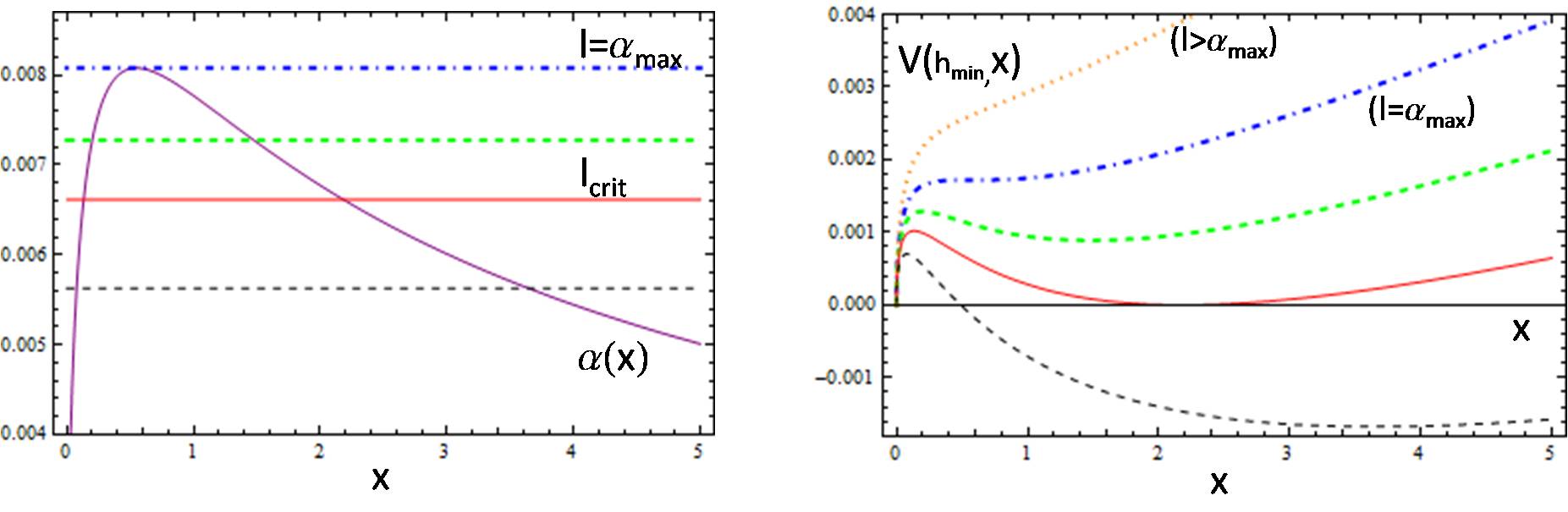}
\caption{\textit{Left}: The quantity $m_0 M^2/\Lambda^3$ define a constant $l=m_0 M^2/\Lambda^3$ whose graph is a straight line. The figure shows a set of lines $l$ intersecting the graph of the $\alpha (x)$ function. \textit{Right:} A set of graphs of the potential associated with each line $l$. The style of the plot of each straight line in the left correspond to that of the style of the potential in the right. The graph of the potential with a continuous (red) style corresponds to a critical straight line $l_c$ determining $x_{min}$ for which $V(x_{min})=0$}.
\label{alphapot}\end{figure}

The second equation in (\ref{dVx}) gives a condition for non-trivial solution(s), if exist. If ${\partial V\over \partial x}=0$, and $\varphi\neq 0$, then the equation $\frac{\partial V}{\partial \varphi}=0$ implies that $\frac{\partial V}{\partial h}=0$, i.e.
\begin{equation}
\frac{1}{A}\frac{\partial V}{\partial h} =-\frac{x}{h^2} +b =0.\label{dVh}
\end{equation}
Solving this expression for $h$, we find a constriction to be satisfied by the set of extremum points:
\begin{equation}
h_{ex}=\sqrt{\frac{x}{b}}.\label{hex}
\end{equation}
This equation represents a curve in the space $x-h$, on which the potential is extremized (i.e. adopts maximum or minimum values). When valuated along this curve, the potential can be seen as a one-variable function (in $x$ or $h$). Substituting (\ref{hex}) in eq. (\ref{Vxh}) we obtain, as a function of $x$:
\begin{equation}
V(x)\vert_{h_{ex}}= A\left[ 2\sqrt{bx}-f(x) \right] \label{Vxhmin}
\end{equation}
In this way, we have seen that the requirement consisting in the two conditions $\frac{\partial V}{\partial \phi}=0$ and $\frac{\partial V}{\partial \varphi}=0$ to be satisfied at the same time, is equivalent (for non-trivial points) to the system of simultaneous equations $\frac{\partial V}{\partial x}=0$, $\frac{\partial V}{\partial h}=0$, which in its turn leads to eq. (\ref{hex}), and to the second equation (\ref{dVx}). To solve this system, we substitute the first one into the last one equated to zero. In this way we obtain the overall condition to be satisfied by the extremum points:
\begin{equation}
\frac{\sqrt{b}}{2}=\sqrt{x_{ex}}\left[1+x_{ex}\log\left(\frac{x_{ex}}{1+x_{ex}}\right)\right],\label{cond1}
\end{equation}
which is a transcendental equation in $x_{ex}$ for any value of $b$, thus can not be solved analytically. Although we can use numerical methods to approach the solution, an explicit algebraic expression for $x_{ex}$ can not be written. This circumstance make difficult to us to know the nature of the extremum points (i.e. whether maximum or minimum), because the standard mathematical procedure, based in calculating the derivatives of second order, ask for such derivatives to be valuated in the suspected extremum points, which are supposed to be previously obtained solving the first order derivatives equations. However, a graphical method could bring a useful alternative, as we will see next:\medskip

Let us translate the equation (\ref{cond1}) as a condition on the parameters of interest, substituting the third equation (\ref{ghb}), and define at a time the function $\alpha$ as follows:
\begin{equation}
\frac{m_0 M^2}{\Lambda^3}=\alpha(x)\equiv \frac{\sqrt{x}}{4\pi^2}\left[1+x\log\left(\frac{x}{1+x}\right)\right]. \label{cond1a}
\end{equation}
We have that $\alpha(x)$ is a bounded function with values between $0\leq \alpha(x) \leq \alpha_{max}$. Let us name $x_0$ the point in which the function $\alpha$ is maximized. This function has
\begin{equation}
\alpha_{max}=\alpha (x_0)\simeq 8.08\times 10^{-3},\quad x_0\simeq 0.55, \label{alphamax}
\end{equation}
and $\alpha(X=0)=0$, increasing monotonically from $x=0$ until $x=x_0$, where the maximum is located. Then, it begins decreasing (also monotonically) to $\alpha\to 0$ as $x$ increases, but never reaching $\alpha=0$ (the $x$ axis is an asymptote for $\alpha(x)$ in the limit $x\to \infty$) (figure \ref{alphapot}, left side).
Now, by assigning specific values to the parameters $m_0$, $M$ and $\Lambda$, a constant $l=m_0 M^2/\Lambda^3$ is determined which graph is a horizontal straight line. Then, we could have three cases: 1) if such line, does not intersect the graph of the function $\alpha(x)$, this means that a solution in $x$ does not exist, i.e. the potential does not possess extremum points. This happens for $l>\alpha_{max}$; 2) if $l=\alpha_{max}$ then there is only one intersection at $x_0$; and 3) if $l<\alpha_{max}$ there are two intersection points, located one before $x_0$, and another after $x_0$, each one representing an extremum point of the potential.\\
This method does not allow us to know whether a given the extremum represents a maximum or minimum, but we can use the information we obtained before: if, according to eq. (\ref{segderivada}), the point $x=0$ is a local minimum, then an extremum located next to it has to be indeed a (local) \textit{maximum}, because there can not be a minimum surrounding another minimum with out existing a maximum between them. Thus, we have $x_{max}<x_0<x_{min}$, therefore $x_{max}<x_{min}$. This is verified in the figures, where we show the minimized potential in $h$, $V(x)\vert_{h_{min}}$, as a function of $x$.\medskip\\

\begin{figure}[h]\centering\includegraphics[scale=.5]{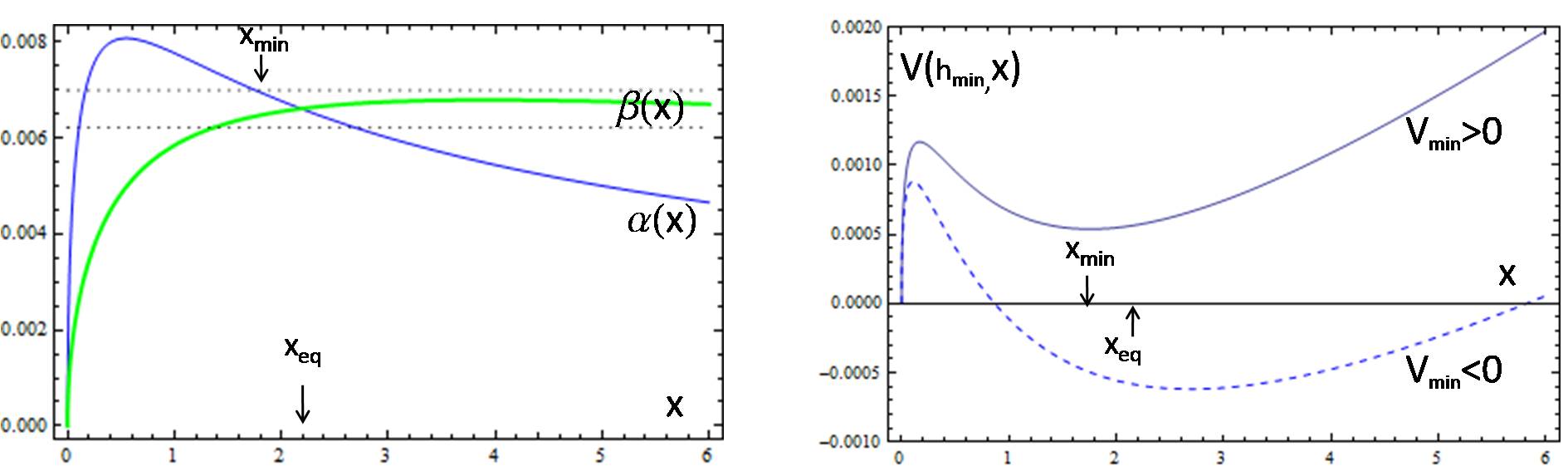}\caption{\textit{Left:} The functions $\alpha(x)$ and $\beta (x)$ intersect at one point $x=x_{eq}$. The horizontal critic straight line $l_{crit}$ is defined by $l_{crit}=\alpha=\beta$. The point $x_{min}$ indicates the value of $x$ in which the potential with $V_{min}>0$ is minimized (right side figure, continuous line style). The point $x_{eq}$ determined by the critical line corresponds to $x_{min}=x_{eq}$ such that $V(x_{min})=0$.}\label{alphabeta1}\end{figure}

Substituting the constraint, equation (\ref{cond1a}), in the expression for the extremized potential, equation (\ref{Vxhmin}), we can write an expression for the potential valuated on the extremum points:
\begin{equation}
V_{ex}=V(x,h)\vert_{(x_{ex},h_{ex})}
={\Lambda^4\over 16\pi^2}\left[\frac{12\pi^2 m_0M^2}{\Lambda^3}\sqrt{x_{ex}}-\log(1+x_{ex})\right].\label{vex}
\end{equation}
Let us take the case for the extremum being a minimum $x_{ex}=x_{min}$. If the minimized potential is positive, i.e. $V_{min}=V(x_{min})\geq 0$, then from equation (\ref{vex}), the $x_{min}$ must satisfy
\begin{equation}
\frac{m_0 M^2}{\Lambda^3} \geq \beta (x_{min})\equiv \frac{1}{12\pi^2}\frac{\log(1+x_{min})}{\sqrt{x_{min}}}, \label{cond12}
\end{equation}
where we have defined the function $\beta(x)$. Therefore, the set of values of $x$ for which the potential is minimized, and at the same time are such that $V_{min}\geq 0$, must satisfy the equations (\ref{cond1a}), and (\ref{cond12}) \textit{simultaneously}. This means that, for these $x$, we have $\alpha\geq \beta$ (figure \ref{alphabeta1}, left side). It is convenient to have a name, let us say $x_{eq}$, for the specific $x$ for which the equality is satisfied, i.e. $\alpha(x_{eq})=\beta(x_{eq})$. Using numerical procedures we can find
\begin{equation}
x_{eq}= 2.18, \quad \alpha(x_{eq})=\beta(x_{eq})=l_c=6.62\times 10^{-3}. \label{alphaeq}
\end{equation}
To help us to visualize how the form of the potential depends on the parameters (remember that $l=m_0M^2/\Lambda^3$), we may build the following picture: beginning with a set of values $m_0$, $M$, $\Lambda$, such that $l>\alpha_{max}$, the minimized potential $V(x)\vert_{h_{min}}$, as a function of $x$, has only a (global) minimum at $x=0$ (the trivial one), and it grows monotonically from there (figure \ref{alphapot}, right side). Lower values of $l$ corresponds to a potential growing slowly as function of $x$. In diminishing the line $l$, it eventually reaches the top of the function $\alpha(x)$, where both graphs intersect at one point $x=x_0$, with $l=\alpha_{max}=\alpha(x_0)$ (figure \ref{alphapot}, left side). On this point $x_0$ (an inflexion point), the potential has a null derivative, and in a way of speaking, it begins to "bend" upside in such a way that it starts acquiring a concavity. From here to further on, for $l<\alpha_{max}$ (the range of definition is until $l=0$), there will be two intersection points between $l$ and $\alpha(x)$, each corresponding to a maximum $x_{max}<x_0$ and a minimum $x_{min}>x_0$. As $l$ is lowering its height, the intersecting point $x_{min}$ slides to the right side, in such a way that $x_{min}$ is increasing (figure \ref{alphapot}). At the same time, because the function (\ref{vex}) decreases with increasing $x$, the minimum of the potential $V_{min}$ (which begins being a positive quantity) is diminishing, and eventually vanishes (i.e. $V_{min}=0$) for certain critical value $l=l_{crit}=\alpha(x_{eq})$ (which is when $\alpha=\beta$). Further on, as $l$ keeps lowering, the minimum $V_{min}$ becomes negative, growing in absolute value as $l$ goes down.\bigskip\\

By using the standard mathematical procedure to analyze the potential, we obtained from the equations for the null first order derivatives, the constraints (\ref{cond1a}) and (\ref{cond12}) on the variables $x,h$. Once we find an extremum point $(x_{ex},h_{ex})$ we can use our definitions to find the correspondent values for the fields $\phi$, $\varphi$ on which the potential is extremized.\\

The total potential (\ref{Vxh}) as a function of the two fields $\phi$, $\varphi$, is written explicitly as
\begin{equation}
V(\phi,\varphi)=\frac{1}{2}m^2\phi^2 +\frac{1}{2}m_0^2\varphi^2
-\frac{\Lambda^4}{16\pi^2}\left[ x+x^2\log\left(\frac{x}{1+x}\right)+\log(1+x)\right] ,\quad x=\frac{m^2\varphi^2\phi^2}{\Lambda^2 M^4}.\label{vexplicit1}
\end{equation}
\begin{figure}[h]\centering\includegraphics[scale=.5]{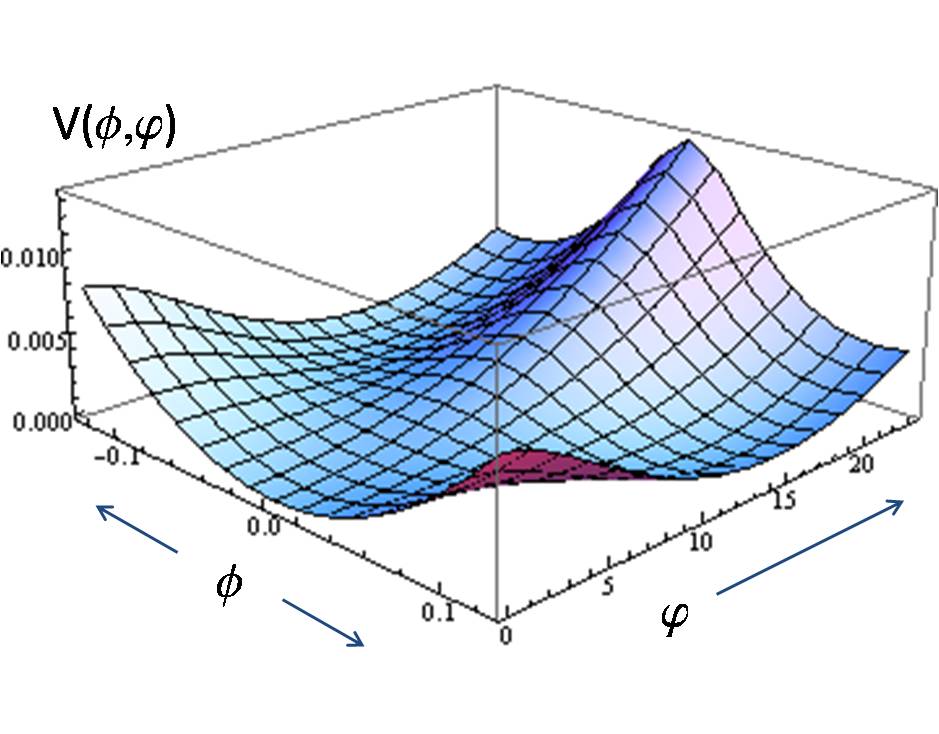}\caption{Total Potential $V=V(\phi,\varphi)$, eq. (\ref{vexplicit1}). As it is a function of two fields, the graph is a two-dimensional surface.}\label{potver02}\end{figure}
By assigning values to the parameters $m_0$, $M$, $\Lambda$ (satisfying $l<\alpha_{max}$) we obtain a fixed value for the extremum $x_{ex}$ through equation (\ref{cond1a}). Substituting in eq. (\ref{hex}) we find $h_{ex}$, which in its turn, can be used together with eq. (\ref{ghb}) to find the extremum value in the field $\varphi$. We have
\begin{equation}
\varphi_{ex}^2=\frac{\Lambda M^2}{m_0}\sqrt{x_{ex}}\label{varphiex}.
\end{equation}
If the extremum refers to the minimum, then the equation above allow us to find $\varphi_{min}$ by making $x_{ex}=x_{min}$. Then we can substitute the now known values $x_{min}$, $\varphi_{min}$ in the definition of $x$, equation (\ref{vexplicit1}), to find also the minimum in the field $\phi$. In this way we obtain
\begin{equation}
\phi_{min}=\frac{M}{m}\sqrt{m_0\Lambda\sqrt{x_{min}}},\quad
\varphi_{min}=M\sqrt{\frac{\Lambda}{m_0}\sqrt{x_{min}}}.\label{camposmin}
\end{equation}
In the case for the extremum $x_{ex}=x_{max}$ being a maximum, we would have analogous expressions for $\phi_{max}$, $\varphi_{max}$. It may be very useful to note the next relation
\begin{equation}
\phi_{min}=\frac{m_0}{m}\varphi_{min}.\label{mm0}
\end{equation}

So we know the relevant points, and we have an idea of how the form of the potential depends on the parameters (figure \ref{potver02}). Now we would like to know the implications for the dynamics of the fields. We must not forget that we are interested in the possibility to implement the inherent phase transition of the NJL model, in the first place. To accomplish this, it is required for the coupling to cover the values $g_{initial}\to g_c \to g_{final}$ successively, or for the field $\varphi_{initial}\to \varphi_c \to \varphi_{final}$. Given that the fields evolve in such a way to make the potential to be minimized, we need the condition
\begin{equation}
V(\varphi_{initial}) >V(\varphi_c) >V(\varphi_{final})\label{vinvfin}
\end{equation}
to be satisfied. In this way the transition could be realized in a natural way, i.e. according to the dynamics of the system, without a "fine tuning" in the initial conditions (for instance in the velocities of the fields; although (\ref{vinvfin}) is not a sufficient condition, as we will see below).\medskip
\begin{figure}[h] \centering \includegraphics[scale=.5]{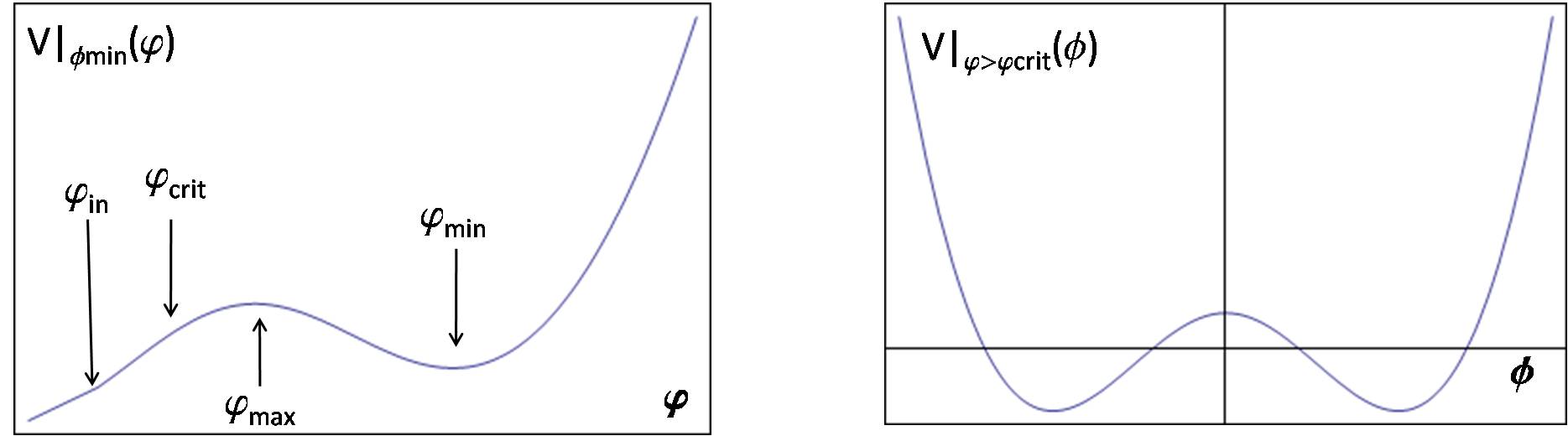} \caption{\textit{Left:} The potential as a function of $\varphi$, minimized in $\phi$. We show how the critical point $\varphi_{c}$ is lesser than the maximum $\varphi_{max}$, according to eq. (\ref{varphigamma}). \textit{Right:} The potential as a function of $\phi$. We show a cross section through $\varphi>\varphi_c$. For a cross section $\varphi<\varphi_c$ it is $V\vert_\varphi\sim \phi^2$.} \label{potver01}\end{figure}
Using the equations (\ref{varphicrit}) and (\ref{varphiex}) we can obtain a relation between the extrema and the critical point:
\begin{equation}
\varphi_{ex}=\gamma\varphi_c ,\quad \mbox{where}\quad \gamma=\frac{1}{1+x\log \left({x\over 1+x} \right) }.\label{varphigamma}
\end{equation}
The coefficient $\gamma$ defined above has $1<\gamma<\infty$, and it is an increasing monotonous function. From this we can see an important result: equation (\ref{varphigamma}) means that $\varphi_c<\varphi_{ex}$, i.e. both extrema are greater than the critical point, so we have in particular $\varphi_c<\varphi_{max}$. In fact, we already know (from the paragraph above equation (\ref{vex})), that the maximum is placed before the minimum, i.e. this is greater than that; thus $\varphi_c<\varphi_{max}<\varphi_{min}$. Now, in order to allow a dynamical phase transition the system must evolve from some initial value $\varphi_i<\varphi_c$, to a final value $\varphi_f>\varphi_c$, with $V(\varphi_i)>V(\varphi_f)$. This can not be achieved, because it should be $V(\varphi_i)>V(\varphi_c)>V(\varphi_{max})$, but we have instead $V(\varphi_{max})>V(\varphi_{any})$ (of course!). This happens, does not matter if $V(\varphi_i)>V(\varphi_{min})$, because $\varphi_{min}$ is beyond $\varphi_{max}$. \footnote{Remember that as time passes, the system is stabilized at the minimum, thus it could (not necessarily) be identified with the final point: $\varphi_{final}=\varphi_{min}$.}
In other words, in going from $\varphi_i$ to $\varphi_{min}$ the system faces a \textit{barrier} of potential with a peak at $\varphi_{max}$ which prevent it to reach the non-trivial minimum, and instead it goes to the trivial one $\varphi=0$ (see figure \ref{potver01}).

\section{Cosmology of the extended NJL model $V=V_I+V_{II}$.}\label{cosmoexnjl}
The study of the universe at an elemental level (without taking into account first order nor higher perturbations) is based in the Friedman -Robertson -Walker -Lemaitre (FRWL) equations, which govern the evolution of the space given the different forms of matter contained in it. If scalar fields compose a part of the content, we must include their own equations of motion (\ref{ecmov1}, \ref{ecmov2}) in addition to the equations for the cosmic background and energy densities of the different components. If we consider contributions from barotropic fluids (equation of state $P_b=\omega_bP_b$ with $\omega_b$ constant), consisting in radiation ($\omega_r=1/3$) and matter ($\omega_m=0$), the complete system of equations to solve is
\begin{equation}
\begin{array}{l} \dot{a} =aH,\\
\\
\dot{H}=-\frac{1}{2}H^2 (4\Omega_r +3\Omega_m) +\dot{\phi}^2+\dot{\varphi}^2,\\
\\
\ddot\phi+3H\dot\phi +\frac{\partial V}{\partial\phi}=0,\\
\\
\ddot\varphi+3H\dot\varphi +\frac{\partial V}{\partial\varphi}=0,\\
\\
\Omega_r ={\Omega_r}_i\left( \frac{{H}}{H_i}\right)^{-2} \left(\frac{a}{a_i}\right)^{-4},\quad
\Omega_m ={\Omega_m}_i\left( \frac{{H}}{H_i}\right)^{-2} \left(\frac{a}{a_i}\right)^{-3}.
\end{array}\label{sistema2}
\end{equation}
For a \textit{flat geometry}, the Friedman equation relating the Hubble parameter $H$ and the total energy density $\rho$ including two scalar fields plus matter and radiation is written
\begin{equation}
H^2={\rho\over 3M_p^2},\quad \mbox{where}\quad \rho=\rho_r +\rho_m +\rho_\phi +\rho_\varphi, \label{H2rho}
\end{equation}
where the Planck mass $M_p=1/\sqrt{8\pi G}$, and the energy densities of the fields $\phi$, and $\varphi$ were defined in the equations (\ref{rhophivarphi1}), (\ref{rhophivarphi2}).
The relative density for the fluid $\alpha$, defined in the usual way is $\Omega_\alpha=\rho_\alpha/(3M_p^2H^2)$, and they must satisfy the following constraint:
\begin{equation}
\Omega_r +\Omega_m +\Omega_\phi +\Omega_\varphi =1.\label{omegatot}
\end{equation}
The well known equation for the acceleration of the scale factor is written in our case as:
\begin{equation}
{\ddot{a}\over a}+{1\over 6M_p^2}\left[  {4\over 3}\rho_r +\rho_m +2(\dot{\phi}^2 +\dot{\varphi}^2-V)\right]=0. \label{ac22}
\end{equation}

Now, we have seen that the potential (\ref{vexplicit1}) has two minima:
\footnote{Strictly speaking, there are five minima: the origin, plus four points corresponding to the non-trivial solution $x=x_{min}$, $h=h_{min}$, because these variables are related in a quadratic way with the fields, in the form $x\sim \phi^2\varphi^2$, $h\sim \varphi^2$. However, because there is a symmetry under a change of sign relating them ($x=x(\pm\phi,\pm\varphi)$, $h=h(\pm\varphi)$), we refer to all these solutions simply as one, "the non-trivial solution".}
one of them is located at the origin $\phi=0$, $\varphi=0$, where $V\vert_{(\phi=0,\varphi=0)}=0$; and the other one associated with the non-trivial solution $x=x_{min}$, $h=h_{min}$ (from where we obtain $\phi_{min}$, $\varphi_{min}$, according to eqs. (\ref{camposmin})) where $V_{min}$ is given by the equation (\ref{vex}) as
\begin{equation}
V_{min}=V(x,h)\vert_{(x_{min},h_{min})} =\frac{4}{3}\Lambda m_0 M^2\sqrt{x_{min}} -{\Lambda^4\over 16\pi^2}\log(1+x_{min}). \label{vtotmin}
\end{equation}
Thus we see that the minimum $V_{min}$ may adopt positive, as well as negative values (or even null).
As well known, in the case of a single scalar field and barotropic fluids evolving in a FRWL universe, the potential is minimized by the field as time passes.\footnote{This fact was also studied before in \cite{njlcosmo} for one field.} In considering two fields we have a similar behavior: the fields in their evolution lead the potential to adopt its minimum value, and the detailed evolution of the fields shall depend on the form of the potential and on the \textit{initial conditions}.\\

Therefore the analysis made for the single field case, is still valid and we have analogous results: depending on the starting point of the fields (in. cond.) we have:\\[3mm]
1.- If the potential is stabilized at a \textit{negative} minimum $V_{min}<0$, then we have a collapsing non-accelerating universe: suppose we start with a positive total energy density $\rho>0$; as the densities of all the other fluids are diminishing, the negative potential $V$ eventually overtakes, making $\rho=0$, i.e. the Hubble parameter\footnote{We have to take a positive initial value $H_i>0$, so that $\dot{a}>0$, i.e. an initially expanding universe; otherwise, if $H_i<0$ the universe would be already contracting. Besides, because the explanation in the main text, the case $H_i>0$ include both expanding and contracting phases, but not viceversa.} $H$ vanishes too (eq. (\ref{H2rho})). This means $H=\dot{a}/a=0$, i.e. $a(t)$ reaches a maximum. Because $H$ is always diminishing (second equation in (\ref{sistema2})) it follows that $H$ becomes negative, meaning $\dot{a}<0$ (first eq. (\ref{sistema2})) i.e. a diminishing scale factor $a(t)$, all the way to $a(t)=0$. From eq. (\ref{ac22}) it can be seen that an acceleration in the neighborhood of the minimum is not possible either, because for this it is required that $(4/3)\rho_r +\rho_m +2(\dot{\phi}^2+\dot{\varphi}^2)<2V$, which clearly never happens as the right side of the inequality is a positive quantity, whereas the left side is negative.\medskip\\

\noindent 2.- If the potential is stabilized at a \textit{vanishing} minimum $V_{min}=0$, the universe expands forever without acceleration, as we now explain:
with an initial value $H_i>0$, the scale factor begins growing, and the total energy density is $\rho=\rho_r +\rho_m +\rho_\phi +\rho_\varphi$. Now, it is known how radiation and matter dilute; we have $\rho_r\sim a^{-4}$, $\rho_m\sim a^{-3}$, so they do not vanish for any finite value of the scale factor, and although decreasing, they remain always positive. On the other hand, the density contribution from the fields is written $\rho_\phi +\rho_\varphi={1\over 2}\dot{\phi}^2+{1\over 2}\dot{\varphi}^2 +V(\phi,\varphi)$, but the kinetic energies are always positive, and even though we may have a vanishing potential $V(\phi,\varphi)\to V_{min}=0$, the sum $\rho_\phi +\rho_\varphi$ remains positive. Thus, the whole sum giving the total $\rho$ never vanishes, and therefore $H$ do not do it either. This means that $\dot{a}\neq 0$, and so the scale factor does not have a maximum value: the universe expands forever. To see the impossibility of the acceleration we use a similar reasoning than before: from eq. (\ref{ac22}), we need $(4/3)\rho_r +\rho_m +2(\dot{\phi}^2+\dot{\varphi}^2)<2V$, but in the neighborhood of the minimum $V\sim 0$, is lesser compared against the never-vanishing $\rho_r$, $\rho_m$.
\medskip\\

\noindent 3.- If the potential is stabilized at a \textit{positive} minimum $V_{min}>0$, then the universe is eventually dominated by the potential, because as time passes, the contributions $\rho_r\sim a^{-4}$, $\rho_m\sim a^{-3}$ to the total energy density $\rho$ in eq. (\ref{H2rho}) are always diminishing (regardless never vanishing) as $a$ is growing, and the kinetic energy do the same as time passes, $E_k=\dot{\phi}^2 +\dot{\varphi}^2\sim 0$.
In the limit of large time the potential overcome and we would have $\rho\sim V_{min}$ with $V_{min}= constant$, so that the potential can be taken for a cosmological constant, which is already known to rise an accelerating universe.\bigskip

As said, the cases 1 ($V_{min}<0$) and 2 ($V_{min}=0$) have been already studied for a single field \cite{}, and the qualitative results are not different in considering two fields instead. If we pretend to explain the observations, among the possibilities described above, the only one which can be realistic is the third one, as the situations 1 and 2 do not include an accelerating phase.\medskip

So we are going to show an example of solution only for the third case. In order to obtain a potential with a positive minimum $V_{min}>0$, we need for the parameters $6.62\times 10^{-3}<m_0M^2/\Lambda^3<8.08\times 10^{-3}$ (equations (\ref{alphamax}, \ref{cond12}, \ref{alphaeq})). Let us take, for instance $\Lambda=m=M$, $m_0=6.95\times 10^{-3}\Lambda$. This define the straight line $l=m_0M^2/\Lambda=6.95\times 10^{-3}$, whose intersections with the function $\alpha(x)$ determine $x_{max}=0.16$, and $x_{min}=1.81$. Then, equations (\ref{varphicrit}, \ref{vex}, \ref{camposmin}) give
\[ \bar{\phi}_{min}=9.67\times 10^{-2},\quad \bar{\varphi}_{max}=7.62,\quad \bar{\varphi}_{min}=13.9\]
\begin{equation}
\bar{V}_{min}=4.67\times 10^{-4},\quad \bar{\varphi}_c =6.28. \label{solsnum}
\end{equation}
The bar over these quantities denotes the \textit{numerical} value.\footnote{To get the real physical value, we need to multiply by certain conversion coefficient. A word on the physical values will be said at the end of this section.}\\
Let us define the initial contribution $\Omega_{nbi}=\Omega_{\phi i} +\Omega_{\varphi i}$ to the content of the universe coming from the fields\footnote{Due to the coefficient $\omega_\alpha$ of the equation of state $P_\alpha=\omega_\alpha\rho_\alpha$ is not a constant for the fields, we say that they form a "non-barotropic" contribution; thus the subindex "nb".} $\phi$, $\varphi$. Consider just for a moment a cosmological constant (here we name it $L$ to avoid confusion with the energy scale $\Lambda$), which has a constant density $\rho_L$, and a matter component with $\rho_m\sim a^{-3}$. We have the ratio
\begin{equation}
{\rho_{m0}\over \rho_{L0}} =\frac{{\Omega_m}_0}{{\Omega_L}_0}=\frac{\Omega_{mi}}{\Omega_{L i}}\frac{1}{({1+z})^3}, \label{OmegaLmat}
\end{equation}
where the subindex "0" denotes, as usual, the present epoch or "today" values.
Now suppose we let the universe to evolve forward in time, starting in the epoch when the amounts of matter and radiation were equal. We know that the redshift for this epoch is $z_{eq}\simeq 3\times 10^3$, with an energy density $\rho_{eq}\sim (1eV)^4$ approx. (in order of magnitude). Also we know the relative densities $\Omega_{m0}\simeq 1/3$ for matter, and $\Omega_{DE0}\simeq 2/3$ for Dark Energy, which we identify with a Cosmological Constant for now. Substituting these values together with $z_{eq}$ in the equation (\ref{OmegaLmat}) we can find that $\Omega_{Li}\simeq 10^{-11}\Omega_{mi}$. Given that our non-barotropic fluid behaves like a cosmological constant near the minimum of the potential, we may have an estimation for $\Omega_{nbi}$, by taking the same order of magnitude than that of $\Omega_{Li}$, i.e. $\Omega_{nbi}\sim 10^{-11}\Omega_{mi}$.
\begin{figure}[h!]\centering\includegraphics[scale=.5]{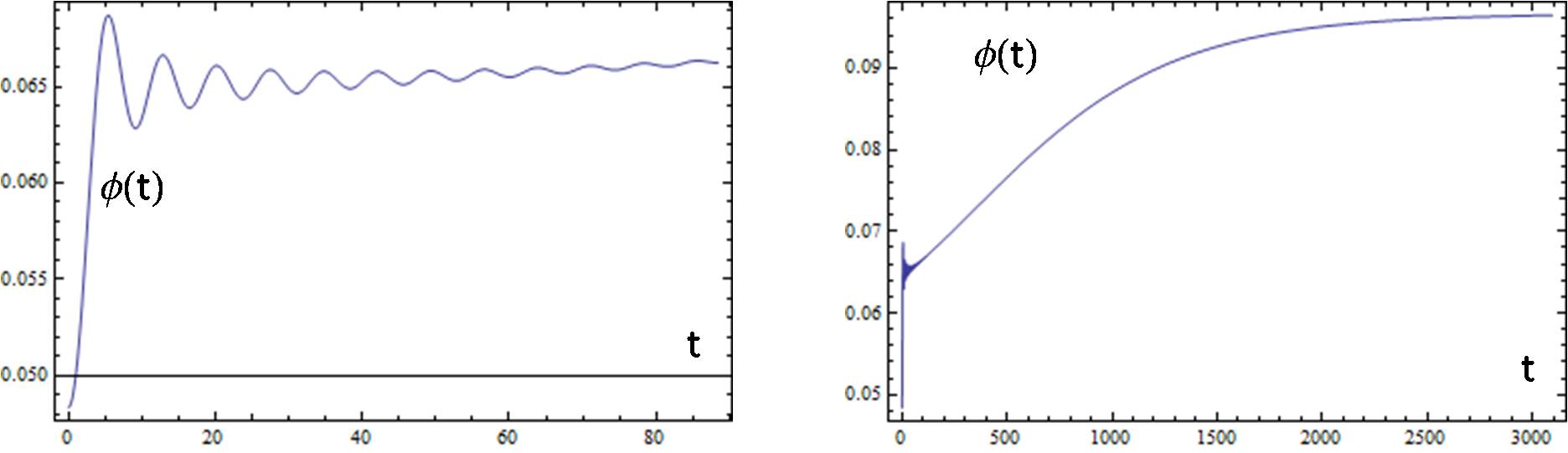}\caption{Evolution in time of the field $\phi$. We show a different time scale in each plot. In the left figure one can recognize an oscillatory behavior.}\label{doscamposphi}\end{figure}
\begin{figure}[h]\centering\includegraphics[scale=.5]{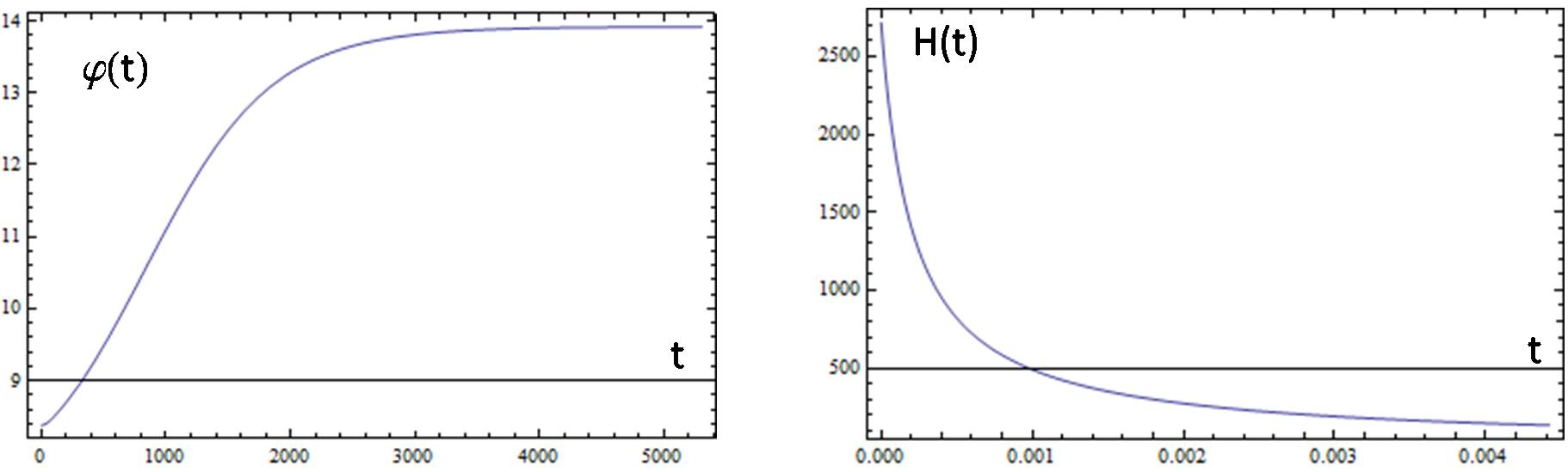}\caption{\textit{Left}: evolution of the field $\varphi$. This field rolls to its minimum without oscillation. \textit{Right:} Hubble parameter $H(t)$. It is decreasing in time, but never becomes to vanish.}\label{doscamposvarphi}\end{figure}
\begin{figure}[h]\centering\includegraphics[scale=.5]{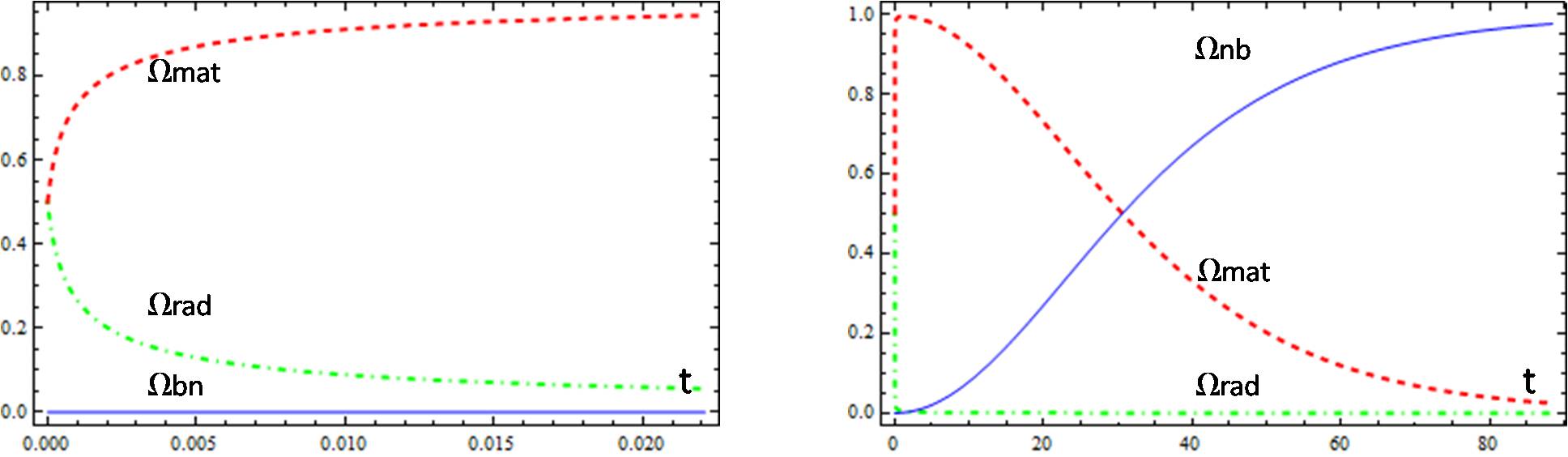}\caption{Relative densities $\Omega_{\alpha}$ for the different involved fluids. The figure in the right side shows a lapse of time $t=2t_0$, which 4,000 times bigger than that of the figure in the left side.}\label{doscamposomega}\end{figure}
\begin{figure}[h]\centering\includegraphics[scale=.5]{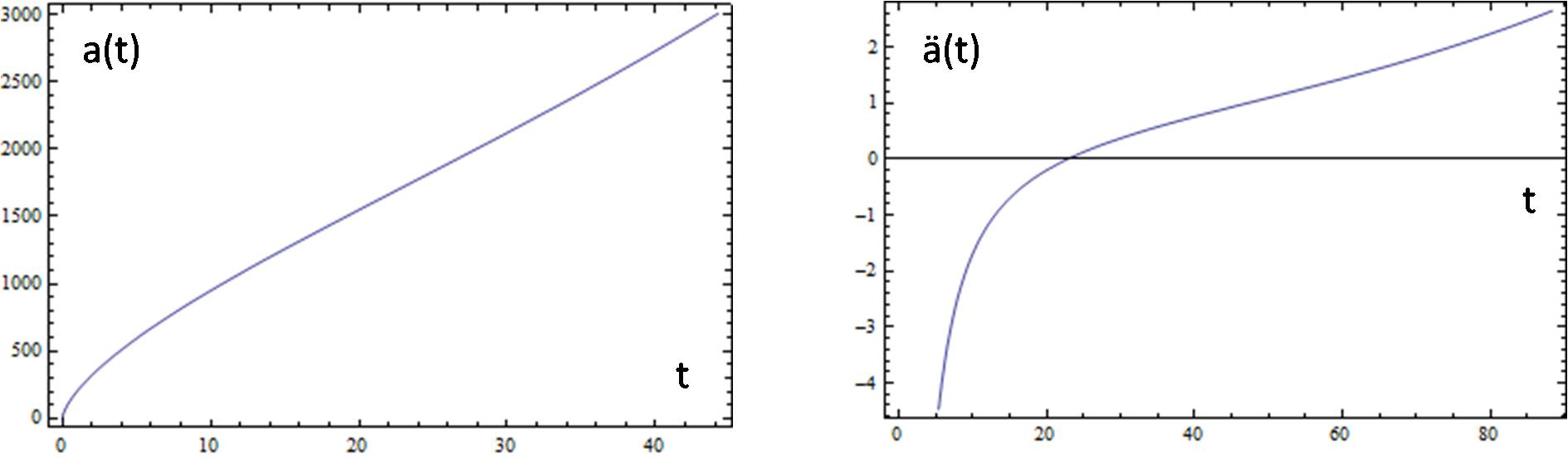}\caption{\textit{Left:} scale factor from $t=0$ to $t=t_0$. \textit{Right:} acceleration of the scale factor from $t=0$ to $t=2t_0$. Here one can observe a period of negative acceleration followed by a period of positive ac., and a transition between them about $t\sim 20$.} \label{doscamposa}\end{figure}
To make the potential stabilize about the non-trivial minimum, we must take $\varphi_i>\varphi_{max}$, with any reasonable value for $\phi$; for instance $\varphi_i=1.1\times \varphi_{max}$, $\phi_i=0.5\times \phi_{min}$. In figures \ref{doscamposphi} and \ref{doscamposvarphi} we see that our predicted values in eq. (\ref{solsnum}) are indeed verified by the numerical solution. We see the field $\phi$ in an oscillatory regime (revealed by zooming about $t=50$) with the mean amplitude evolving to the minimum $\phi_{min}\sim 0.097$, whereas the field $\varphi$ goes to reach its own minimum $\varphi\sim 13.9$. Due to the fact that each field has its own term ${\partial V\over \partial \phi}$, or ${\partial V\over \partial \varphi}$ in the equation of motion, they react to the potential in a different way. Thus, in this specific case the field $\varphi$ is over damped and do not oscillate in its way to the minimum.

The relative densities $\Omega_\alpha$ evolve as expected: radiation dilutes faster than matter, and due to having taken a very small initial density $\Omega_{nbi}$, we have to wait long time to see it dominate (figure \ref{doscamposomega}). Due to this same fact, the scale factor is allowed to grow without accelerating up to a size $a_0/a_i=(1+z_{eq})\sim 3000$ (figure \ref{doscamposa}), which is a number compatible with the one from the standard cosmological model in considering the evolution of the universe since the epoch of matter-radiation equality. In the figure, an accelerating period is seen to begin about $t=t_{ac}=20$, where $\ddot{a}=0$.\medskip\\
Perhaps it would be worth to clarify that, the moment $t_{ac}$ at which the acceleration begins has not to coincide necessarily with the moment $t_{DE}$, $\Omega_{DE}(t_{DE})=\Omega_m (t_{DE}) +\Omega_r (t_{DE})$ at which the Dark Energy begins to dominate. For our solution in particular we obtain the following numbers:
\begin{center}
\begin{tabular}{|c|c|c|}
\hline $a_0/a_i=3006$ & $z_0=0$ & $t_0=1$ \\ \hline
$a_{DE}/a_i=2157$ & $z_{DE}=0.39$ & $t_{DE}=0.69\times t_0$ \\ \hline
$a_{ac}/a_i=1713$ & $z_{ac}=0.76$ & $t_{ac}=0.52\times t_0$ \\ \hline
\end{tabular} \end{center}
Notice that the acceleration period begins \textit{before} Dark Energy becomes dominant, $t_{ac}<t_{DE}$. One may check that the same fact results in considering a Cosmological Constant instead of our scalar fields model.\medskip\\

Let us now estimate some real physical values. The total energy density today is about $\rho_0=E_0^4\sim (10^{-3}eV)^4$, and the Dark Energy contribution is $\rho_{DE0}=\Omega_{DE0}\rho_0$. If we identify our non-barotropic NJL fluid with DE, we would have $\rho_{nb}=\rho_{DE0}$. Now, in the limit of stabilized fields about the minimum\footnote{We must keep in mind that in general, the fields could be in a rolling regime, so the kinetic energies would not be negligible. Therefore we must be careful in the conditions that we are talking about.} the energy density of our NJL fluid is $\rho_{nb}=V_{min}$. Then, (approximating $\Omega_{DE0}\simeq 2/3\sim 1$) we may write $\rho_{DE0} =E_0^4 =V_{min}$, and using the result obtained in equation (\ref{vtotmin}) we find
\begin{equation}
E_0^4=
\frac{3}{4}\Lambda^4 \left[ \sqrt{x}\frac{m_0 M^2}{\Lambda^3}-\frac{1}{12\pi^2} \log (1+x) \right].
\end{equation}
This can be expressed in a suitable form with the help of the equation (\ref{cond1a}), and defining a coefficient $\epsilon$ as
\begin{equation}
E_0^4=\epsilon\Lambda^4,\quad
\epsilon=\frac{3}{4}\sqrt{x}\left[\alpha(x)-\beta(x) \right].\label{E0ep}
\end{equation}
The more the quantity $m_0M^2/\Lambda^3$ gets close to $\alpha_{eq}$, the more smaller becomes the coefficient $\epsilon$. So, the equations (\ref{cond1a}) and (\ref{E0ep}) with $E_0\simeq 10^{-3}eV$ are constraints on the parameters $\Lambda$, $M$, $m_0$. A discussion of the possibilities in this respect are not relevant for the time being, and will be left for another study.

\section{Summary and Conclusions.}\label{conclu}
The fermion theory NJL gives place to two physical phases, each arising depending on the coupling strength. In a previous paper \cite{njlcosmo}, each phase was seen to have different (and interesting) implications in considering a NJL fluid as a material component of the universe. To take quantum corrections into account, an effective potential was calculated, and written in terms of an auxiliary scalar field $\phi$, providing an useful equivalent description of the system. However, as the coupling is not a dynamical variable, each phase had to be treated one at once. In the present work, we argued that, by introducing a dynamical field dependent coupling $g$  in terms of additional scalar field $\varphi$ we may obtain a dynamical cosmological phase transition. In order to such transition to be realized, the field $\varphi$ must pass through a critical value $\varphi_c$ along its evolution. To investigate the new two-field physical system, we studied a new two-dimensional potential, eq. (\ref{vexplicit1}), and defined some useful auxiliary functions $\alpha$ eq. (\ref{cond1a}), $\beta$ eq. (\ref{cond12}). We have seen that the potential may, or may not have a non-trivial minimum (i.e. other than the origin $\phi=0$, $\varphi=0$, where $V_{min}(trivial)=0$ for \textit{every cases}) depending on the set of parameters $m_0$, $M$, $\Lambda$ defining the relevant parameter $l=m_0M^2/\Lambda^3$. If $l>\alpha_{max}$, then the potential has only one minimum at the origin; if on the other hand  $\alpha=\beta=l_c<l<\alpha_{max}$  then the potential has two minima (the origin and the non-trivial related to $x_{min}$) with the non-trivial $V_{min}>0$; and finally if $0<l<\alpha=\beta=l_c$, then the potential has also two minima (the origin and the non-trivial), with $V_{min}<0$ (see figures \ref{alphabeta1}, \ref{potver02}). \medskip

Now, according to the current observations, at present time the universe is in an accelerating period,  caused  \textit{Dark Energy}, emulated to a good extent by a Cosmological Constant. Thus, we wonder whether our model can be useful to be compatible with this fact.

If the  potential has  $l>\alpha_{max}$ (i.e. no minima other than the origin) a phase transition can certainly be realized, given that the fields minimize the potential, it would be enough to take an initial value $\varphi_i>\varphi_c$. In this way the system would evolve from a strong coupling $g>g_c$ (fermion condensate $\phi\neq 0$), to a weak coupling $g<g_c$ (massless fermions $\phi=0$), to finally stabilize at the origin. However, in this case there would be no acceleration, unless we choose such a large initial value of $\varphi_i$ to make the condition $(4/3)\rho_{ri} +\rho_{mi} +2(\dot{\phi_i}^2+\dot{\varphi_i}^2)<2V_i$ from eq. (\ref{ac22}) be satisfied. This, would be an "early" acceleration, because it arises before the system stabilizes on the minimum, and before allowing the matter and radiation components to dominate the universe and dilute to the present day values. So, this case is not useful as a possible model of Dark Energy.\footnote{It arises the question of the viability of this model in the context of the Early Cosmic Inflation. However, we will not go further to study this possibility right now.}\medskip

On the other hand, for a potential with $l<\alpha_{max}$ (thus having two minima, at the origin and  at a the non-trivial value) we found that the relevant points are distributed as $\varphi_c<\varphi_{max}<\varphi_{min}$.
Therefore, if the initial amplitude is $\varphi_i<\varphi_c$, the field cannot evolve to an "expected" value $\varphi_e >\varphi_c$, because $V(\varphi_i)<V(\varphi_e)$, and the fields minimize the potential. Even though one could still choose an initial amplitude in the form $\varphi_c<\varphi_i<\varphi_{max}$, in such a way that a phase transition could be realized, but in the opposite direction, from strong to weak coupling, this procedure however is not acceptable, because it would imply a fine tuning in the initial condition $\varphi_i$, besides not including an accelerating period.

The only alternative left is that the initial amplitude begins in a region which could make the field stabilize in the non-trivial minimum. This requires $\varphi_{max}<\varphi_i$, but then the field would be disabled to roll to $\varphi_c$. Being so, as time increases the densities of matter and radiation decrease and the fields stabilize. With $V_{min}>0$ the potential ends up dominating the total energy density and behaves like a cosmological constant, making the universe to accelerate. In this way (adjusting the initial amplitude $\varphi$, and choosing the parameters to get the right potential $V_{min}>0$), we could have a viable model of Dark Energy, but at the price of avoid the phase transition.\\

Although the choosing could be seen as a defect of the model, this would be on the same footing than other physical theories, where one has to take the "right" parameters to fit the observations. Moreover, from the philosophical point of view it is perhaps more appealing to postulate a QFT particle than a non-understood kind of constant energy.\\

Finally, we must not forget that when the potential has a vanishing minimum $V_{min}$, the fields dilute like $\rho_\alpha\sim a^{-3}$, so the
model can be seen as describing a matter component (Dark Matter?), and therefore still could offer a useful theory which may be worth to study further.

\section*{Acknowledgment}
{\small A.M.  acknowledges financial support from  UNAM PAPIIT Project No. IN101415}

\thebibliography{}

\footnotesize{

\bibitem{SN.1}
Perlmutter et al., Astrophys. J. 517, 565 (1999); A. G.
Riess et al., Astron. J. 116, 1009 (1998); R. Amanullah et al., Astrophys. J. 716, 712 (2010),

\bibitem{SN.2} C. Contreras, et al. Astron.J. 139 (2010) 519-539 arXiv:0910.3330,
M. Hicken et al. Astrophys.J. 700 (2009) 331-357 arXiv:0901.4787,
S. Jha,  et al. Astron.J. 131 (2006) 527-554 astro-ph/0509234,
A. G. Riess,  et al. Astrophys.J. 659 (2007) 98-121 astro-ph/0611572 46455850950,
N. Suzuki, et al. Astrophys.J. 746 (2012) 85 arXiv:1105.3470,

\bibitem{CMB} A. D. Sakharov.  1966.  Sov.Phys.JETP,22,241,
P.J.E. Peebles,  et al. Astrophys.J. 162 (1970) 815-836,
R.A. Sunyaev,  et al. Astrophys.Space Sci. 7 (1970) 3-19,

\bibitem{wmap9}  WMAP Collaboration (Bennett, C.L. et al.) Astrophys.J.Suppl. 208 (2013) 20 arXiv:1212.5225 [astro-ph.CO]

 \bib{planck}
  P.~A.~R.~Ade {\it et al.} [Planck Collaboration],
  arXiv:1502.01589 [astro-ph.CO],
  P.~A.~R.~Ade {\it et al.} [Planck Collaboration],
  arXiv:1502.01590 [astro-ph.CO].
  P.~A.~R.~Ade {\it et al.}  [ Planck Collaboration],
  arXiv:1303.5076 [astro-ph.CO].

\bibitem{LSS.1}
B. A. Reid et al., Mon. Not. R. Astron. Soc. 404, 60(2010);
W. J. Percival et al., Mon. Not. R. Astron. Soc. 327, 1297(2001);
M.Tegmark \textit{et al.},  Mon.Not.Roy.Astron.Soc. 328 (2001) 1039 ,
D. G.York,  et al. Astron.J. 120 (2000) 1579-1587

\bibitem{LSS.2}
Padmanabhan, Nikhil et al. Mon.Not.Roy.Astron.Soc. 427 (2012) 3, 2132-2145 arXiv:1202.0090,
Michael J. et al. Mon.Not.Roy.Astron.Soc. 401 (2010) 1429-1452,
D.H. Jones,  et al. Mon.Not.Roy.Astron.Soc. 399 (2009) 683 arXiv:0903.5451,
K. S.Dawson,  et al., Astron.J. 145 (2013) 10 arXiv:1208.0022,
L. Anderson,  et al., Mon.Not.Roy.Astron.Soc. 441 (2014) 24-62 arXiv:1312.4877,
T. Delubac,  et al., Astron.Astrophys. 574 (2015) A59 arXiv:1404.1801

\bibitem{BAO.1} D. J. Eisenstein, et al.Astrophys.J. 633 (2005) 560-574
\bibitem{BAO.2} R. Tojeiro, et al. Mon.Not.Roy.Astron.Soc. 440 (2014) 2222 arXiv:1401.1768,
Isabelle et al. Astron.Astrophys. 563 (2014) A54 arXiv:1311.4870,


\bibitem{NJL}
Y.~Nambu and G.~Jona-Lasinio,
  Phys.\ Rev.\  {\bf 122}, 345 (1961),
 Y.~Nambu and G.~Jona-Lasinio,
  Phys.\ Rev.\  {\bf 124}, 246 (1961).

\bibitem{njlcosmo}
L. Quintanar and  A. de la Macorra arXiv:1511.06210

\bibitem{NJL.ax}
  A.~de la Macorra and G.~G.~Ross,
  Nucl.\ Phys.\ B {\bf 404}, 321 (1993);
  A.~de la Macorra and G.~G.~Ross,
  Nucl.\ Phys.\ B {\bf 443}, 127 (1995).


\bibitem{massgap} W. A. Bardeen, C. T. Hill, and M.Lindner
Phys. Rev. D41 (1990) 1647,   M.Lindner and D. Ross, Nucl. Phys. (1992) 30

\bibitem{NJLInflation}T.Inagaki., S.D. Odintsov, H.Sakamoto arXiv:1509.03738

\bibitem{DE.rev}
E. J. Copeland, M. Sami,
and S. Tsujikawa, Int. J. Mod. Phys. D 15, 1753 (2006).

\bibitem{SF.Peebles}
B.~Ratra and P.~J.~E.~Peebles,
  Phys.\ Rev.\  D {\bf 37}, 3406 (1988),C.~Wetterich,
  Astron.\ Astrophys.\  {\bf 301}, 321 (1995)
  [arXiv:hep-th/9408025].

\bibitem{tracker}
P.J. Steinhardt, L. Wang, I. Zlatev  Phys.Rev.Lett. 82 (1999) 896, arXiv:astro-ph/9807002;
Phys.Rev.D 59(1999) 123504, arXiv:astro-ph/9812313
9.

\bibitem{quint.ax}
  A.~de la Macorra and G.~Piccinelli,
  Phys.\ Rev.\  D {\bf 61}, 123503 (2000)
  [arXiv:hep-ph/9909459];
  A.~de la Macorra and C.~Stephan-Otto,
  Phys.\ Rev.\  D {\bf 65}, 083520 (2002)
  [arXiv:astro-ph/0110460].

\bib{DEparam}

M. Doran and G. Robbers, J. Cosmol. Astropart. Phys. 06 (2006) 026;
E.V. Linder, Astropart. Phys. 26, 16 (2006);
D. Rubin et al. , Astrophys. J. 695, 391 (2009) [arXiv:0807.1108];
J. Sollerman et al. , Astrophys. J. 703, 1374 (2009)[arXiv:0908.4276];
M.J. Mortonson, W. Hu, D. Huterer, Phys. Rev. D 81, 063007 (2010)  [arXiv:0912.3816];
S. Hannestad, E. Mortsell JCAP 0409 (2004) 001 [astro-ph/0407259];
H.K.Jassal, J.S.Bagla, T.Padmanabhan, Mon.Not.Roy.Astron.Soc. 356, L11-L16 (2005);
S. Lee,  Phys.Rev.D71, 123528 (2005)

\bib{DEEoS.ax}
A. de la Macorra, arXiv: 1511.04439

\bibitem{GDE.ax}
  A.~de la Macorra,
  Phys.\ Rev.\  D {\bf 72}, 043508 (2005)
  [arXiv:astro-ph/0409523];
  A.~De la Macorra,
  JHEP {\bf 0301}, 033 (2003)
  [arXiv:hep-ph/0111292];
   A.~de la Macorra and C.~Stephan-Otto,
  Phys.\ Rev.\ Lett.\  {\bf 87}, 271301 (2001)
  [arXiv:astro-ph/0106316];

\bibitem{GDM.ax}
  A.~de la Macorra,
  Phys.\ Lett.\  B {\bf 585}, 17 (2004)
  [arXiv:astro-ph/0212275],A.~de la Macorra,
  Astropart.\ Phys.\  {\bf 33}, 195 (2010)
  [arXiv:0908.0571 [astro-ph.CO]].

\bibitem{IDE}
 S.~Das, P.~S.~Corasaniti and J.~Khoury,Phys.  Rev. D { 73},
083509 (2006), arXiv:astro-ph/0510628;
A.~de la Macorra,  
 Phys.Rev.D76, 027301 (2007), arXiv:astro-ph/0701635

\bibitem{IDE.ax}
  A.~de la Macorra,
  JCAP {\bf 0801}, 030 (2008)
  [arXiv:astro-ph/0703702];
  A.~de la Macorra,
  Astropart.\ Phys.\  {\bf 28}, 196 (2007)
  [arXiv:astro-ph/0702239].

}

\end{document}